# Crossover from one- to three-dimensional behavior in the S = ½ Heisenberg antiferromagnet $Cu(N_2H_5)_2(SO_4)_2$

A. Paduan-Filho, A. P. Vieira, J. G. A. Ramon and R. S. Freitas*

Instituto de Física, Universidade de São Paulo, 05314-970 São Paulo, SP, Brazil.

*Corresponding author: freitas@if.usp.br

From experimental and theoretical analyses of magnetic and specific-heat properties, we present the complete magnetic phase diagram of the quasi-one-dimensional antiferromagnet $Cu(N_2H_5)_2(SO_4)_2$. On cooling and at zero magnetic field this compound enters a one-dimensional regime with short-range magnetic correlations, marked by a broad maximum in the specific heat and magnetic susceptibility at $T_{max} \sim 2$ K, followed by a tridimensional antiferromagnetically ordered phase below $T_N \sim 1$ K induced by small interchain couplings. The intermediate-temperature one-dimensional regime can be modeled using exact quantum-transfer-matrix calculations, which offer a compatible description of the nonmonotonic behavior of $T_{max}$ as a function of the magnetic field, giving $J = 4.25$ K for the intrachain exchange parameter. The analysis of magnetic specific-heat and susceptibility data at low temperature indicates that the interchain exchange couplings are an order of magnitude smaller than the coupling inside the chains.

# I. INTRODUCTION

Quasi-one-dimensional magnetic materials are systems in which exchange couplings are much stronger along one particular direction, and thus can be described in terms of weakly interacting chains of atoms. These systems are quite interesting from both experimental and theoretical viewpoints, since they usually exhibit a dimensional crossover, as the temperature is lowered or the magnetic field is increased, between one-dimensional and three-dimensional behavior.[1,2,3,4,5,6] In the particular case of insulating spin-1/2 isotropic systems with antiferromagnetic intrachain couplings, the magnetic degrees of freedom can be described, except at very low temperatures, by the one-dimensional Heisenberg model with nearest-neighbor interactions, whose Hamiltonian is given by

$$\mathcal{H} = J \sum_{i=1}^{N-1} \vec{S}_j \cdot \vec{S}_{j+1} - g\mu_B H \sum_{j=1}^{N} S_j^z, \qquad (1)$$

in which $\vec{S}_j = \left(S_j^x, S_j^y, S_j^z\right)$ is a spin-1/2 object sitting on the $j$th of $N$ sites along the chain, $J > 0$ is an antiferromagnetic exchange coupling, $g$ is the gyromagnetic factor, $\mu_B$ is the Bohr magneton, and $H$ is the applied magnetic field. This model can be exactly solved, exhibits a gapless spin excitation spectrum, and does not allow for long-range magnetic ordering, although its ground state is critical, being characterized by quasi-long-range order, with power-law decay of spin-pair correlations[7]. However, at sufficiently low temperatures, materials in which the weak interchain couplings are nonfrustrating usually exhibit three-dimensional magnetic ordering, associated with the opening of a gap in the excitation spectrum. Measurements in these compounds with applied magnetic fields can be performed down to very low temperatures, allowing both the comparison between experimental and theoretical results and a detailed investigation of the dimensional crossover and the phase transition. This feature, together with the possibility of synthesizing a variety of inorganic copper-based ($Cu^{2+}$) materials showing quasi-one-dimensional S=1/2 structures, makes the investigation of this class of compounds very attractive.

The compound copper-hydrazine sulphate, $Cu(N_2H_5)_2(SO_4)_2$, has the crystal structure displayed in Fig. 1, which is formed by Cu ions surrounded by four $SO_4$ groups and two $N_2H_5$ groups.[8] The atomic arrangement in this system determines a chain bridged by the $SO_4$ groups, with the chains being mutually linked via the $N_2H_5$ ligands. Along the chains the direct exchange interactions should be of minor importance since the distance between adjacent Cu ions, 0.5402 nm, is rather large. Although the separation between the chains is only slightly larger than the intra-chain distance between the metal ions, the more complicated superexchange pathways between the chains are expected to allow an approximate description of this compound in terms of isolated magnetic linear chains, with only weak interchain couplings. Likewise, as a result of the possible pathways for the magnetic interactions and structural features of this compound, a low-anisotropy magnetic behavior is expected, suggesting a description in terms of the one-dimensional Heisenberg model down to a temperature scale set by the weak interchain couplings. Possible corrections due to easy-plane XXZ-like or Dzyaloshinksi-Moriya interactions could in principle be investigated using single crystals rather than the polycrystalline samples currently at our disposal. This question therefore remains open.

Early investigations of the magnetic properties of this compound focused on magnetization ($M$) and specific heat measurements down to temperatures of 1.2 K.[8,9] Results for the susceptibility $\chi$ (approximated as $M/H$ at $H = 1$ T) at high temperatures $T$ are well fitted by a Curie-Weis law, $\chi = A/(T - \Theta_W)$, yielding $\Theta_W \simeq -3$ K, a value corresponding to an antiferromagnetic interaction between the Cu ions. Measurements of the ac susceptibility at low fields suggest the existence of a rounded maximum around $T \sim 2$ K (Ref. 8), compatible with the behavior predicted from Eq. (1) with $J = 4$ K and $g = 2.12$. (Notice that our definition of $J$ differs by a factor of 2 from that in Ref. [8].) Data for the magnetic specific heat measured at zero field, which displays a broad maximum at T = 1.95 K, can be likewise fitted by theoretical calculations based on the Heisenberg model with an exchange coupling $J = 3.98$ K. These early data were used as benchmarks for our experiments.

In this work we extend the experimental study of the quasi-1D antiferromagnet $Cu(N_2H_5)_2(SO_4)_2$ to temperatures down to 0.2 K, allowing us to uncover effects of interchain couplings only faintly hinted at by the higher-temperature measurements of Refs. [8] and [9]. Systematic magnetic and thermal measurements were performed, focusing on the magnetization and specific heat for different values of the magnetic field. Special attention was given to the low-temperature region in which features associated with interchain interactions turn out to be more pronounced. We find that, as the temperature is lowered, there is a dimensional crossover from the 1D behavior previously investigated to a fully 3D behavior characterized by antiferromagnetic long-range order.

The paper is organized as follows. In Sec. II we describe the experimental apparatus and the measurement protocol. An analysis of the experimental results and a theoretical description of the 1D regime, along with estimates for the interchain couplings, are presented in Sec. III. Finally, in Sec. IV we summarize the paper and present our conclusions.

## II. EXPERIMENT

We worked with polycrystalline samples of Cu(N$_2$H$_5$)$_2$(SO$_4$)$_2$ prepared by adding aqueous solutions of CuSO$_4$.H$_2$O to an aqueous solution of N$_2$H$_6$SO$_4$, as described previously.[10] X-ray analysis confirmed the single-phase triclinic structure reported in Ref. [11].

The magnetization was measured in magnetic fields up to 15 T and in temperatures down to 0.6 K using a vibrating sample magnetometer (VSM) in a He$^3$ refrigerator and at low fields and high temperatures (above 2 K) with a SQUID magnetometer (Quantum Design MPMS). The specific-heat data were obtained using a Quantum Design Dynacool system with a dilution refrigerator option, under several applied magnetic fields up to 9 T, down to the temperature of 0.2 K. The addendum heat capacity was measured separately and subtracted.

## III. ANALYSIS AND DISCUSSION

Figure 2 shows the measured specific heat under several applied magnetic fields in the temperature range between 0.2 and 20 K. Under zero magnetic field, two features are clearly visible, a broad maximum centered at $T_{max} \simeq 2$ K and a peak at $T_N \simeq 1$ K. Temperatures and amplitudes of both features evolve independently as the field is varied, as will be discussed below. The Debye lattice contribution to the specific heat can be estimated using a fit of the high temperature part of the measured specific heat to an asymptotically series of odd powers of the temperature, $C_{latt}(T) = aT^3 + bT^5 + cT^7$. This procedure is usually adopted for insulating magnetic systems,[12] and in our case produces the curve shown in Fig. 2. After the subtraction of the lattice contribution, the magnetic specific heat, $c_m(T)$, can be obtained as shown in Fig. 3 for $H = 0$ and $H = 6$ T. The Cu$^{2+}$ nuclear contribution to the specific heat is negligible at the temperatures investigated here.[13] The entropy change associated with the magnetic degrees of freedom can be estimated by integrating the magnetic specific heat, $\Delta S_m = \int \frac{c_m}{T} dT$. The result is shown in Fig. 4 and is in very close agreement with the expected value for an $S = ½$ system, namely, $\Delta S_m = R \ln 2$. As shown in Fig. 3, the high-temperature maximum at $T_{max}$ becomes rounded and shorter, while shifting to higher temperatures at high magnetic fields. At the same time, the low-temperature peak shows the exactly opposite behavior. As anticipated in the work by Klaaijsen et al.,[9] the maximum at $T_{max}$ is related to the development of short-range magnetic correlations along the chains. We associate the lower transition temperature $T_N$ with the appearance of a 3D long-range order induced by interchain interactions. The inset of Fig. 4 shows the low-temperature part of the magnetic heat capacity, which has the $T^3$ behavior characteristic of antiferromagnetic magnons.[14,15,16]

In order to investigate the 1D magnetic regime, corresponding to $T \gtrsim 2$ K, we resort to exact quantum-transfer-matrix (QTM) calculations[17,18] for the Hamiltonian described by Eq. (1). Explicitly, this method takes advantage of the well-known Bethe ansatz equations for the Heisenberg model and of the Trotter decomposition to write the magnetic free energy per spin $f(T,h)$ as the infinite summation

$$f(T,h) = -T\ln 2 - \frac{T}{2}\sum_{l=1}^{\infty} \ln\left\{\frac{(1+p_l^2)(1+\bar{p}_l^2)}{\left[4\pi T\left(l-\frac{1}{2}\right)/J\right]^4}\right\},$$

in which $h = g\mu_B H$ and the complex conjugate numbers $p_l$ and $\bar{p}_l$ satisfy

$$p_l = -\frac{4\pi T}{J}\left(l-\frac{1}{2}\right) - \frac{2ih}{J} + \frac{p_l}{1+p_l^2} + \frac{2iT}{J}\sum_{j=1}^{\infty}\ln[L(p_l,p_j)L(p_l,-\bar{p}_j)],$$

with $L(x,y)$ a complex function given by

$$L(x,y) = \frac{1-(1-ix)^{-1}+iy}{1-(1+ix)^{-1}+iy}.$$

Cutting off the numerical summation at high values of $l$ we obtain very accurate values of the free energy. Thermodynamic properties are then calculated from the usual recipe of the canonical ensemble (see e.g. Ref. [19])

From fits of the zero-field magnetic specific heat data, shown in Fig. 3, we obtain $J = 4.25$ K as an estimate for the intrachain exchange coupling. In the presence of an applied field, both specific-heat and magnetization measurements (to be detailed below) can be well fitted in the 1D regime (i.e. for temperatures $T \gtrsim 2.5$ K) with a gyromagnetic factor $g = 2.2$. These estimates for $J$ and $g$ are in good agreement with those from earlier work.[8]

Figure 5 shows curves of magnetization versus field for temperatures between $T = 0.6$ K and $T = 4.12$ K. Notice that theoretical curves closely follow the experimental data for temperatures above ~ 2.5 K. The inflection point present in the curve for $T = 0.6$ K is consistent with both quantum Monte Carlo simulations and experimental data for low-dimensional antiferromagnetic Heisenberg systems (see e.g. references [20] and [21]). Figure 6 shows experimental data for the temperature dependence of the susceptibility $\chi$ (estimated as $M/H$ at $H = 1$ T). As the temperature is lowered, the susceptibility first exhibits a broad maximum around $T = 2.2$ K, characteristic of a 1D system with only short-range magnetic correlations, then passes an inflection point and starts to increase again, as shown in Fig. 6. Similar behavior is also observed in other low dimensional magnets.[22,23] The increase of the susceptibility below the inflection point is a typical property of quasi-1D systems, which can be understood from the Fisher relation,[24] which establishes a correspondence between a derivative of $T\chi$ and the zero-field magnetic specific heat. The same relation allows a check of the value $T_N$ of the transition temperature obtained from the magnetic specific-heat data. As shown in the inset of Fig. 6, a peak in the derivative of $T\chi$ indeed occurs at $T_N \simeq 1$ K.

A theoretical description of the 3D transition is hindered by the fact that the 3D Heisenberg model has not been solved exactly. However, a number of estimates for the order of magnitude of the interchain couplings can be obtained provided we make the approximation of assuming a tetragonal rather than triclinic crystalline structure for our material, which amounts to representing interchain couplings by a single number $J'$.

The relation between the transition temperature $T_N$ and the interchain coupling $J'$ can be approximately established as follows. A combination of mean-field and field-theoretical calculations by Schulz[25] suggests

$$|J'| \simeq \frac{T_N}{1.28\sqrt{\ln\left(\frac{5.8J}{T_N}\right)}},$$

yielding for our data $|J'| \simeq 0.44$ K. On the other hand, an empirical formula based on quantum Monte Carlo calculations[26] reads

$$|J'| \simeq \frac{T_N}{0.932\sqrt{\ln\left(\frac{2.6J}{T_N}\right) + \frac{1}{2}\ln\ln\left(\frac{2.6J}{T_N}\right)}},$$

yielding $|J'| \simeq 0.64$ K.

Yet another estimate of $J'$ can be obtained from the low-temperature magnetic specific heat curve. A linear spin-wave calculation[27] for a tetragonal antiferromagnet under zero magnetic field leads to a low-temperature effective Hamiltonian

$$H = E_0 + \sum_{\vec{k}} \hbar\omega_{\vec{k}}\left(\alpha^\dagger_{\vec{k}}\alpha_{\vec{k}} + \beta^\dagger_{\vec{k}}\beta_{\vec{k}}\right),$$

in which $E_0$ is a constant, the $\alpha$s and $\beta$s are bosonic operators and the frequencies $\omega_{\vec{k}}$ are related to the wavevectors $\vec{k}$ by the dispersion relation

$$\hbar\omega_{\vec{k}} = (J + 2J')\sqrt{1 - \gamma_{\vec{k}}^2},$$

in which

$$\gamma_{\vec{k}} = \frac{J\cos(k_x a) + J'[\cos(k_y b) + \cos(k_z b)]}{J + 2J'}.$$

At low temperatures, the main contribution to thermodynamic properties comes from the long-wavelength (small $|\vec{k}|$) limit. A standard calculation of the bosonic partition function (see e.g. Ref. [19]) then leads to a specific heat

$$c_m(T) = AT^3, \qquad A = \frac{4\pi^2 k_B N_A}{15J^2 J'}.$$

From the data in the inset of Fig. 4 we obtain $A = 4.26$ J/K$^4$mol, yielding $J' = 0.28$ K.

The previous estimates, obtained by different approximations, although differing by a factor 2, consistently point to interchain couplings which are an order of magnitude weaker than the intrachain couplings $J$. Further support to this conclusion is provided by the fact that the magnetization curve at $T = 0.6$ K (see Fig. 5) leads to an estimate of the zero-temperature saturation field between 6.5 T and 7 T, a value which is around 15% above the value $H_{sat} = 2J/g\mu_B$ valid for the strictly 1D Heisenberg antiferromagnetic model.

Figure 7 summarizes in a field-versus-temperature diagram the various regimes of magnetic behavior, as extracted from our results. Down triangles and stars indicate the positions of the maxima in the magnetic specific heat according to experimental data and 1D QTM calculations, respectively. These points can be loosely identified with a "boundary" separating the higher-temperature isolated-spin regime (marked IS-PM in Fig. 7) from a one-dimensional regime characterized by short-range magnetic correlations (marked 1D-PM). The nonmonotonic aspect of the curve defined by the maxima in the specific heat has its origins in the competition between the antiferromagnetic intrachain coupling $J$ and the external field $H$, which set the scale for the energy excitations in the paramagnetic phase. Upon cooling at very small fields, antiferromagnetic short-range correlations develop at temperatures of order $J$. Increasing the external field weakens the antiferromagnetic correlations, shifting the maximum in the specific heat towards lower temperatures. At very high fields, the intrachain coupling becomes irrelevant, and the maximum in the specific heat turn out to be associated with thermal fluctuations deviating the spins from the field direction, requiring temperatures whose scale is set by $H$. This gives rise to the linear behavior of the boundary on the upper right in Fig. 7. Together, the isolated-spin regime and the one-dimensional regime define the paramagnetic phase, which is separated from the antiferromagnetic phase (marked 3D-AF in Fig. 7) by a genuine phase boundary indicated by up triangles and diamonds. Notice the reentrant behavior of the 3D-AF boundary, indicating that, for a small range of temperatures, one can go from the 1D to the 3D and back to the 1D behavior by increasing the magnetic field starting from 0 T. This feature is also a characteristic of quasi-1D systems.[28]

## IV. SUMARY AND CONCLUSIONS

In summary, we have used two different experimental techniques, specific-heat and magnetization measurements, to fully characterize the magnetic interactions and construct the phase diagram of the quasi-one-dimensional compound $Cu(N_2H_5)_2(SO_4)_2$. Our results confirm that this magnetic system follows the predicted magnetic and thermal behavior deduced from the theoretical approach used for the high temperature regime ($T \gtrsim 1$ K) where 1D short-range correlations dominate. The observed specific-heat maximum in this temperature region has a characteristic nonmonotonic dependence on the applied magnetic field. At low temperatures and for fields up to 7 T a transition is very clear in the specific heat and magnetic susceptibility, signaling the development of the long-range antiferromagnetic order. The phase boundary separating 1D from 3D behavior is fully determined, showing a curvature at low fields that attests to the low-dimensional character of the magnetic units in this system.

## ACKNOWLEDGMENTS

RSF and AP-F would like to acknowledge support from the Brazilian agencies CNPq (478031/2013-0 and 302880/2013-5, respectively) and FAPESP (2015/16191-5). APV thanks financial support from NAP-FCX and J.G.A.R. thanks the CAPES for sponsorship.

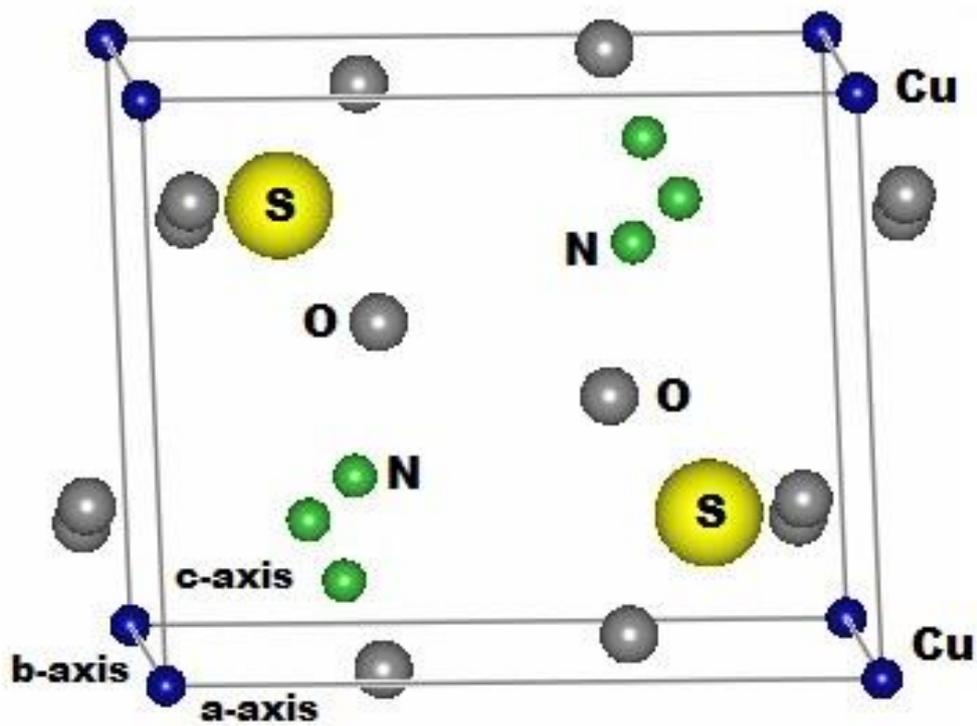

Figure 1 – The triclinic structure of $(N_2H_5)_2Cu(SO_4)_2$. The Cu ions linear chains run along the direction of the *b*-axis.

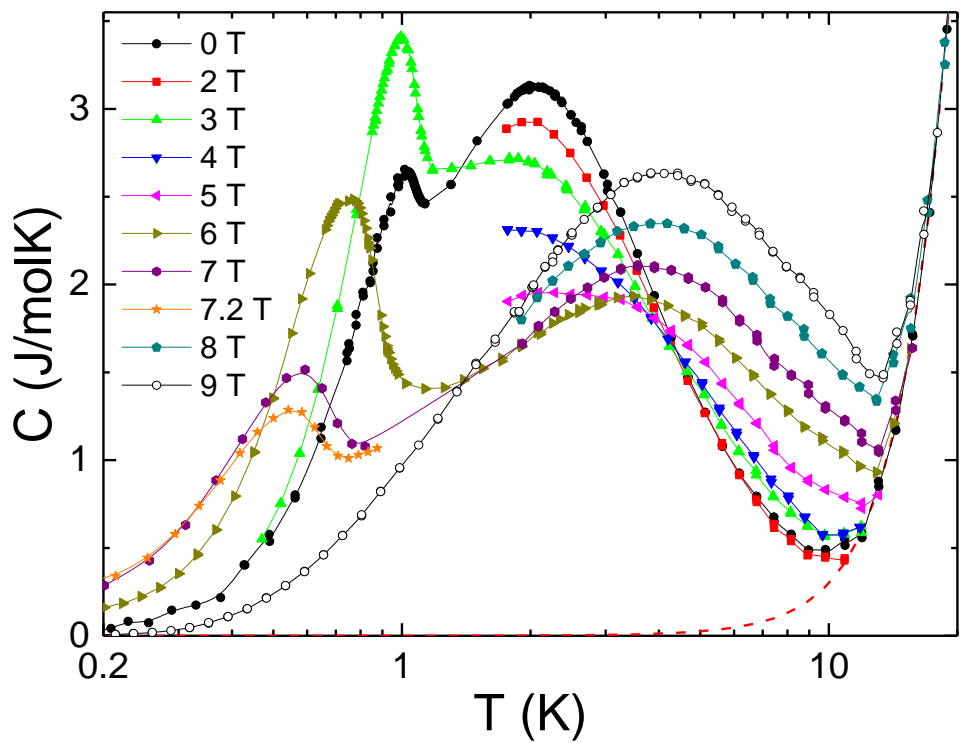

Figure 2 – Specific heat of $Cu(N_2H_5)_2(SO_4)_2$ as a function of the temperature for several applied magnet field values. The dashed red curve represents the lattice contribution which is dominant above T = 12 K. The continuous lines are only guides to the eyes.

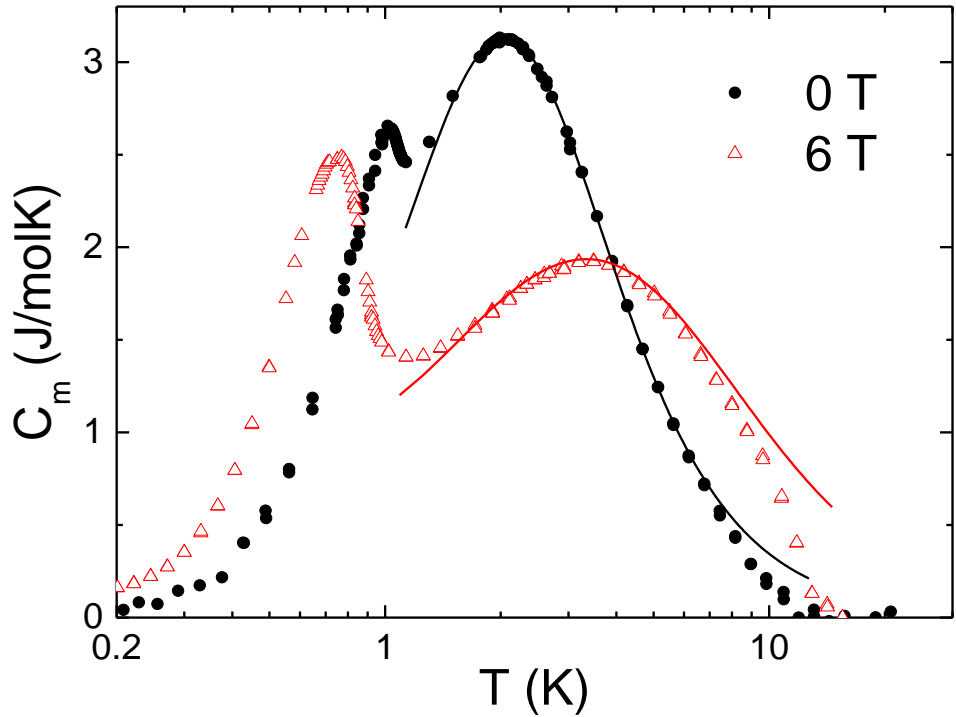

Figure 3 – Temperature dependence of the magnetic specific heat measured under $H = 0$ and $H = 6$ T. The solid lines represent high temperature fits based on theoretical calculations as explained in the main text. Discrepancies between experimental and theoretical results above $T \simeq 10$ K are due to inadequacies of the low-temperature polynomial expansion of the lattice contribution which is subtracted from experimental data to yield the magnetic specific heat.

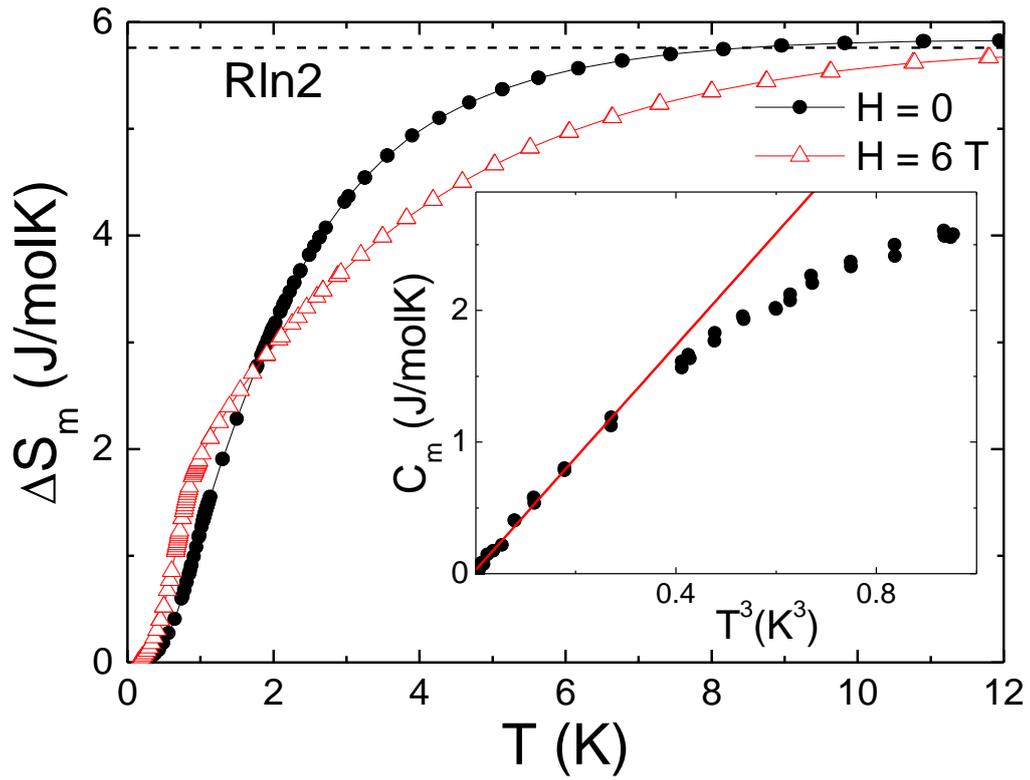

Figure 4 – Integrated magnetic entropy obtained after the subtraction of the lattice contribution to the specific heat. The insert shows the expected $T^3$ behavior of the magnetic specific heat for antiferromagnetic magnons at low temperatures.

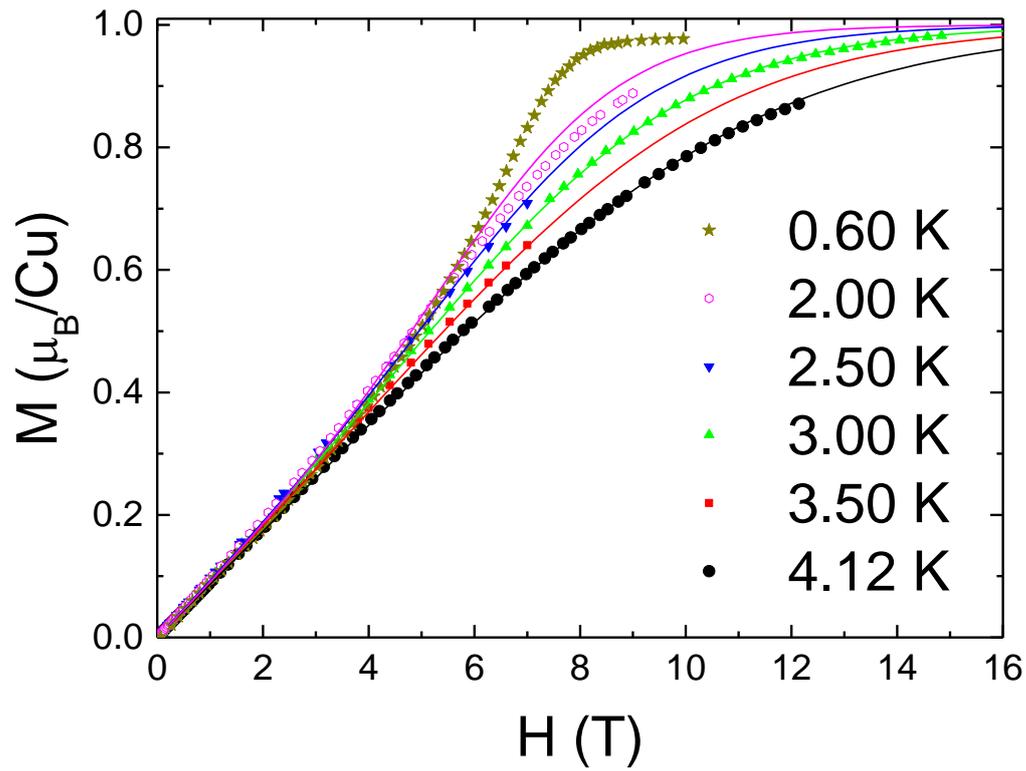

Figure 5 - Magnetization as a function of field at various temperatures. The solid lines are the theoretical curves assuming a linear chain model as described in the main text.

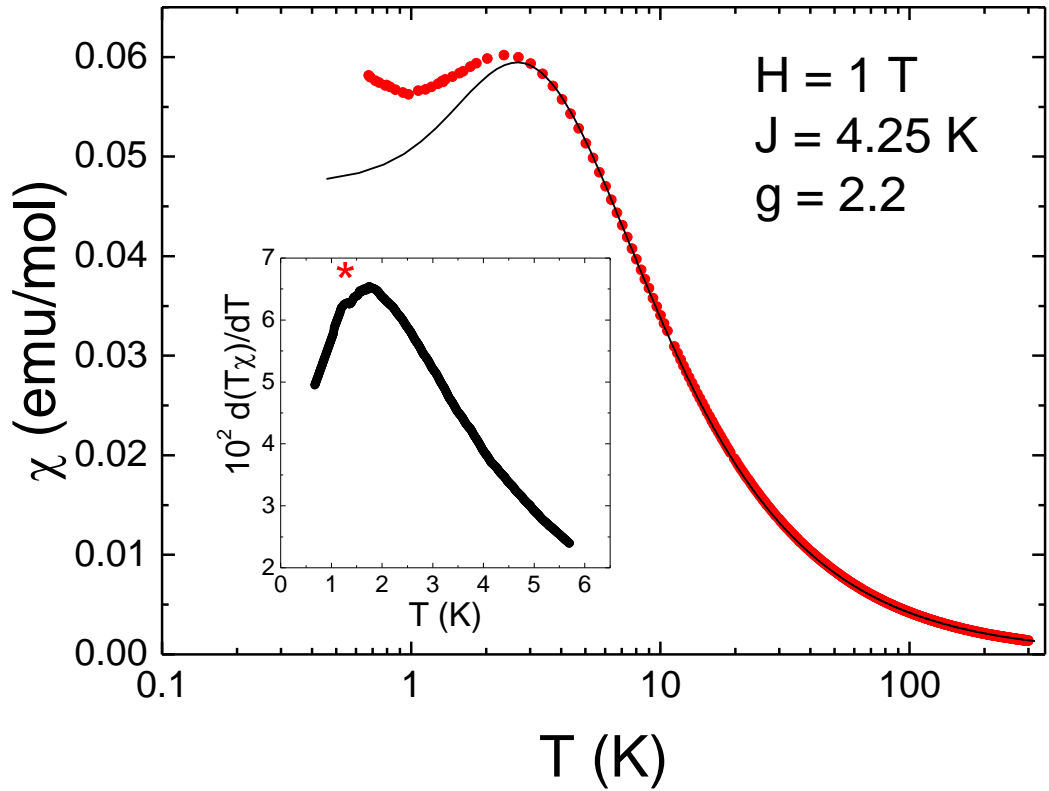

Figure 6 - Temperature dependence of the magnetic susceptibility in a field $H = 1$ T. The solid line corresponds to the theoretical curve for the 1D Heisenberg model with parameters $J = 4.25$ K and $g = 2.2$. The inset shows the derivative of the susceptibility with the indicated critical temperature associated with the 3D long-range ordering transition according to the Fisher's relation.[24]

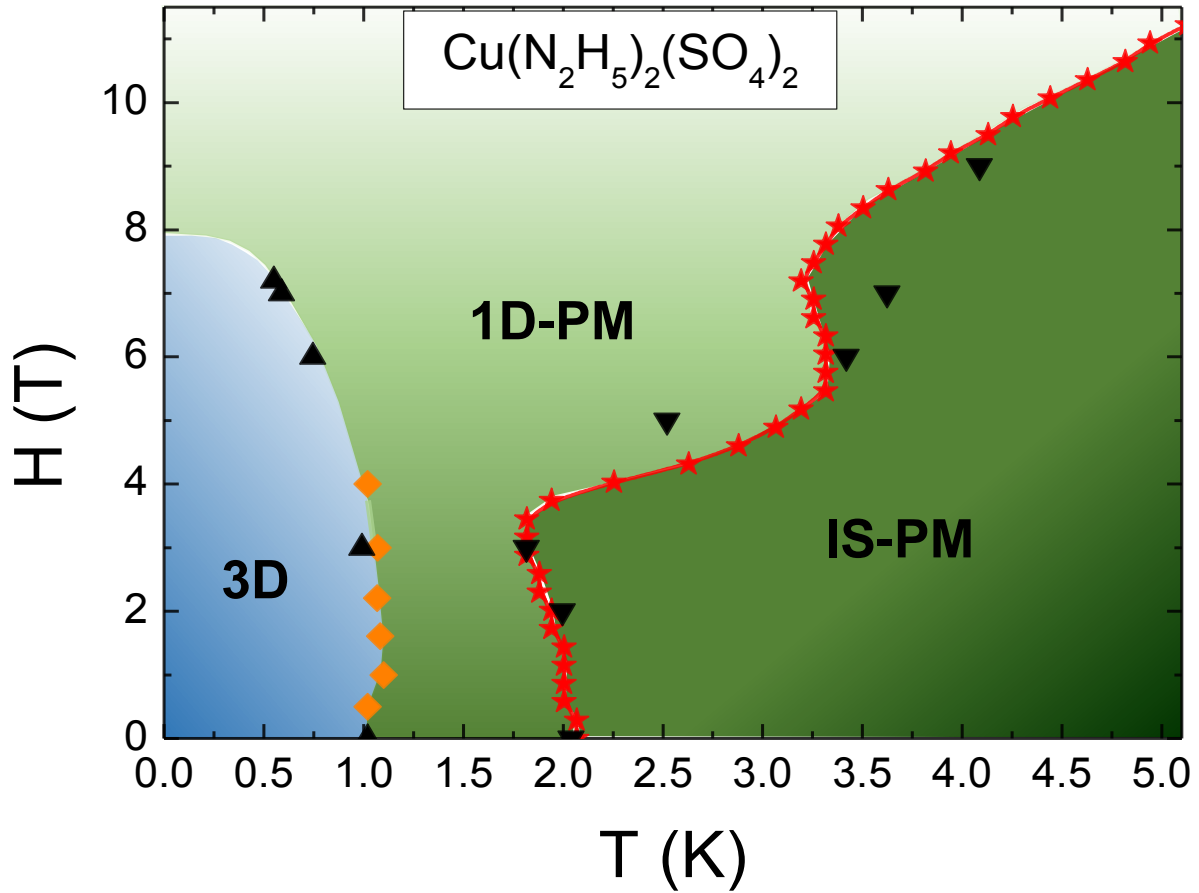

Figure 7 - The magnetic phase diagram of Cu(N$_2$H$_5$)$_2$(SO$_4$)$_2$ showing the paramagnetic phase (PM) and the antiferromagnetic phase (AF). The paramagnetic phase consists of two regions, the one-dimensional paramagnetic regime (1D-PM) and the isolated-spin paramagnetic regimes (IS-PM), loosely separated by a curve defined by the temperatures at which, for a given field, a broad maximum occurs at the magnetic specific-heat curve. At low temperatures and fields, a fully-ordered antiferromagnetic state develops (3D-AF), separated from the paramagnetic phase by a genuine transition line. Up triangles indicate the positions of narrow peaks in the measured specific heat. Diamond symbols correspond to narrow peaks in the derivative of the magnetic susceptibility, as shown in the inset of Fig. 5. Down triangles correspond to the positions of broad maxima in the measured magnetic specific heat and the stars are the corresponding calculated values using the 1D model.